\documentclass[epj]{svjour}
\usepackage{graphics}
\begin{document}
\title{Pulsar: repeatable Lagrangian singularity}
\author{Alexander V. Chigirinsky\inst{1}
\thanks{\emph{Present address:} Br. Trofimovyh st., 4/2, 101, 
Dnepropetrovsk, 49068, Ukraine, tel: 380(56)3708437}%
}                     
\offprints{A.V. Chigirinsky}          
line
\institute{private person, \email{chigirinsk@ua.fm}}
\date{Received: date / Revised version: date}
%
\abstract{
In general, the interior of radially symmetric self-gravitating sphere 
is considered in terms of hydrostatic equilibrium (HSE). This approach 
implies the possibility of the static being of a body. Such a static 
state is assumed to be the result of asymptotic damping of the process 
of formation. It is shown here that the damping of this process is 
impossible: if a sphere vibrates radially, then compressional wave is 
singular at the centre; dynamical singularity has no intermediate stages 
of the fading; the HSE-state is unachievable. Self-gravitating sphere 
perpetually vibrates in essentially singular way, it contains dynamical 
central region -- pulsatile Lagrangian cavity. Theoretical properties of 
this cavity indicate that this is a pulsar. A pulsar is common 
structural feature for every self-gravitating structure. 
} 
\maketitle
\section{Introduction}
\label{intro}
Modern understanding of the interior of a spherical non-rotating 
heavenly body relies upon three theoretical pillars: law of gravity; 
idea of the hydrostatic equilibrium (HSE); speculative equation of the 
state of a matter. Static stability of a heavenly body is seen as the 
equilibrium of two opposite factors - the squeezing gravity and the 
pressure which prevents the collapse of a body. Classic HSE-idea is 
expressed as the differential equation $\partial p/\partial r$=$-\rho 
g$, where $p$, $\rho$, $g$ are pressure, mass density and gravitational 
acceleration correspondingly. There are at least three well-known 
explanations of this equation: (i)~the Archimedian equilibrium of some 
differential frame of a matter inside a sphere but at non-central 
position; (ii)~the special case of the Euler-Lagrange equation of 
hydrodynamic flow at zero velocities; (iii)~the condition of the energy 
stationarity of the sphere \cite{zel81}. All these approaches presume 
\emph{a priori} the possibility of the ideal static state of a sphere. 
Such an ideal HSE-state is assumed to be basic one, and then it is 
assumed that radial vibration of a sphere could be expressed in terms of 
the finite perturbations. And vice versa, as far as radial vibrations 
are expected to be finite, these can subside somehow (\emph{e.g.} due to 
viscosity) and sphere goes to the quiescent HSE-state. This standpoint 
is as much valid as all the functions involved are non-singular, since 
the idea of the dynamical singularity and the idea of the finite static 
state are mutually exclusive notions.
 
The question is that centre of an ideal sphere is essentially peculiar 
point. In every respect, central point must be considered as a degraded 
sphere when radially symmetric motion is examined. How does the spheric 
wave propagates through it? The centre of a solid sphere is motionless 
and absolutely rigid point due to symmetry.\footnote{Laplacian wave 
equation of a solid ball yields trivial finite solution. Non-trivial 
finite solution is non-interpretable one -- central point oscillates 
alternating in sign; its positive displacement could be interpreted as 
the appearance of the cavity (\emph{sic}); its negative displacement has 
no reasonable meaning. Usually, the spherical wave is associated with 
the Dirac delta function -- \emph{i.e.} with the wittingly singular 
structure.} Then, the propagation of the single finite-energy wave-front 
through the degraded sphere could be associated with instantaneous 
infinite density. However, then continuous vibrational spectrum results 
in the permanent presence of a wave at the centre, hence -- in the 
permanently singular central density. \emph{E.g.}, the collapse of a 
protostellar cloud could be interpreted as the falling phase of the 
compressional wave. Extensive analytical and numerical examinations of 
this process in terms of the Euler-Lagrange equation (\emph{e.g.} 
\cite{bre98}) give either the dispersion of the initial cloud or the 
creation of central `core' of singular density. Intuitively clear that: 
(i) permanent singular  density is inadmissible; (ii) even momentary 
singular density can not be modeled by the continuous matter -- the 
presence of a wave at the centre results in the catastrophic disruption 
of a matter. The imminence of such a disruption gives another idea of a 
wave propagation -- it could be associated with the geometric pulsation 
of the cavity. Due to this cavity, a sphere falls inside of itself, 
self-collides, and bounces off. This cavity is not a theoretical 
discovery -- this one is an inseparable element of the accurate 
Lagrangian definition of the mass configuration `radius via mass' 
$r(m)$.

\section{Lagrangian singularity}
\label{seq:ls}
\textbf{Definitions}~Variable Lagrangian sphere of the radius $r(m)$ 
contains invariable mass $m$; $r(0)$$\geq$0. Strictly monotonous 
increasing function $r(m)$ maps (0,$M$)$\leftrightarrow$($r(0),r(M)$), 
where $M$ is the total mass of the system. Evidently, every sphere 
$x$:$x$$\leq$$r(0)$ contains naught, and every sphere 
$y$$:$$y$$\geq$$r(M)$ contains the total mass $M$. To let single-valued 
mapping, two boundary points must be excluded from the range of the 
definition. Hereafter, boundary values of a function $f$ are interpreted 
as $f(0)$=$\lim\limits_{m\rightarrow+0}f(m)$; 
$f(M)$=$\lim\limits_{m\rightarrow M-0}f(m)$. Let an increment $df$ be 
associated with $dm$; the function $\rho$=$(dv/dm)^{-1}$ is set as the 
mass density, where $dv$ is the volume of the frame $dm$. Formally, 
Lagrangian sphere can be defined equivalently either in terms of $v(m)$ 
or $s(m)$: $s$=$4\pi r^2$. Virtual displacement $\delta r$ corresponds 
to the deformation of the sphere $r$ into $r+\delta r$; the perturbation 
$\delta f$ is caused by $\delta r$. Let the perturbation $\delta f$ be 
also associated with its temporal duration $\delta t$: $\delta 
f$=$\dot{f}\delta t$. The Lagrangian definition in form of the identity 
$\delta m\equiv 0$ is the continuity equation (CE). \textbf{End of 
definitions}

Lagrangian singularity arises immediately with the notion `virtual 
deformation of the degraded sphere $r(0)$=0'. This variation must be 
considered according to the theoretical routine.  Lagrangian variation 
$\delta r(0)$ of the central point $r(0)$=0 gives rise to the 
\emph{finite} sphere $r'(0)$=$\delta r(0)$. This is the act of the 
creation of a cavity from naught, it can not be expressed in terms of 
the differential relations. In particular, the relation $\delta v$=4$\pi 
r^2\delta r$ is valid on the condition $|\delta r|$$\ll$$r$. Hence, the 
adoption of this relation induces $\delta r$=0 at $r(0)$=0, and we have 
hidden forbidenness of the virtual variation of the sphere $\delta r(0)$ 
at the state $r(0)$=0. The matter of the paradox is that the theoretical 
analysis is unclosed until $\delta r(0)$$>$0 is examined. Further, the 
relation $\dot{v}$=4$\pi r^2\dot{r}$ is evidently singular: finite 
volumetric velocity $\dot{v}$ of the collapse corresponds to infinite 
linear velocity $\dot{r}$ at $r$=0. Thus $\delta r(0)$$>$0 can not be 
examined in terms of \emph{a priori} quasistatic variation at $r(0)$=0. 

\textbf{Euler-Lagrange equation} of hydrodynamic flow of radially 
symmetric structure\footnote{This is formal modification of the 
HSE-proof \cite{zel81}. Kinetic energy and non-trivial virtual 
displacement $\delta r(0)$ are taken into consideration.}  becomes 
evidently singular in Lagrangian terms. Let self-gravitating spheric 
structure be self-contained system. Then, the variance $\delta r$ is 
permissible if it does not effect upon the total energy of the system 
$E$:
\begin{eqnarray}
{\delta E=0=\int_0^M\delta (dK)+\int_0^M\delta (dH)+\int_0^M\delta 
(dU),}\nonumber
\end{eqnarray}
where $dK$, $dH$, $dU$ are kinetic, heat, and gravitational energies of 
the frame $dm$ correspondingly. Theirs variations are definable 
functions 
\[
\delta (dK)=\delta (\frac{1}{2} {\dot{r}}^2~dm)
=\dot{r}\ddot{r}\delta t~dm=\ddot{r}~dm~\delta r,\\
\]
\[
\delta (dU)=g(r)~dm~\delta r=G\frac{mdm}{r^2}~\delta r,
\]
and, assuming adiabatic deformation of the sphere,
\[
\delta (dH)=-p\delta(dv)=-pd(\delta v)=-d(p\delta v)+\delta 
v~\frac{dp}{dm}~dm,
\]
where $p(m)$ is the pressure. The separate integration of the last 
variation yields 
\[{
\delta H= -(p\ \delta v)|_0^M + \int _0^{~M} \frac{dp}{dm} \delta v dm 
}\]
The equity 
$(p\delta v)|_{m=0} - (p\delta v)|_{m=M}$=0
is the requirement, since the self-consistent system is examined. This 
is the work done by two environmental pressures: $p(M)$=0 is the 
pressure of the Universal vacuum; $p(0)$=0 meets the definition of an 
evacuated cavity $r(0)$.\footnote{In the original work \cite{zel81}, the 
equity $(p\delta v)|_{m=0}$=0 is explained as following: $(p~4\pi 
r^2\delta r)|_{m=0}$=0, since $r(0)$=0. Differential relations $\delta 
v$=$4\pi r^2\delta r$, $\delta (dv)$=$d(\delta v)$, and $\delta 
(1/r)$=$-\delta r/r^2$ are valid at the condition $|\delta r|$$\ll$$r$. 
Hence $\delta r(0)$=0, and the HSE-equation is not proven at the point 
$r(0)$=0 in this case.}\\
The model can be rescaled to the dimensionless parameters and functions 
$m$=$M\mu$, $t$=$T_0t'$, $r$=$R_0r'$, $p$=$p_0p'$ with the scaling 
factors $R_0$, $p_0$, $T_0$ which obey the relations
\[{
GM^2/R_0=p_0R_0^3=MR_0^2/T_0^2
}\]
Then, by the relations $\delta r$=$\dot{r}\delta t$ and $\dot{v}$=$4\pi 
r^2\dot{r}$ (and by omitting all the symbols prime), 
\begin{eqnarray}
0\equiv 
\frac{\delta E}{\delta t}=\int _0^{~1} (\ddot{r} + 4\pi 
r^2\frac{dp}{d\mu} + \frac {\mu}{r^2})~\dot{r}~d\mu 
\label{sing}\end{eqnarray}
As the velocities $\dot{r}$ are assumed be arbitrary, the simplified 
Euler-Lagrange equation
\begin{eqnarray}
\ddot{r} = - 4\pi r^2\frac{dp}{d\mu} - \frac {\mu}{r^2}\label{sele}
\end{eqnarray}
meets the eq.\ref{sing} on two boundary conditions $p(0)$=$p(1)$=0. 
Obviously, the condition $\dot{r}$$\equiv$0 is tautological to the 
requirement $\dot{E}$$\equiv$0, hence static state of a body is 
self-sufficient and incognizable notion, since the expression in 
parenthesis eq.\ref{sing} is arbitrary in this case: the HSE-equation 
can not be considered as the particular case of the eq.\ref{sele} at 
$\dot{r}$$\equiv$0. At some reasonable initial conditions
\begin{eqnarray}
r(\mu, t_i)=r_i(\mu);~r(0, t_i)>0;~\dot{r}( \mu, 
t_i)=u_i(\mu)\label{init}
\end{eqnarray}
and the constrain function $p$=$p(\rho)$, a heavenly body almost 
permanently exists `on the fly' due to infinite set $ \{t_j 
\}$:~$r(0,t_j)$=0 of self-collisions. Key concept of the present 
approach consists in the relation $r(0, t_i)>0$: finite initial 
conditions can be specified at the presence of the finite initial 
Lagrangian cavity only.\footnote{Graphically, this is an analogy of the 
evident idea that non-trivial two-body problem can not be formulated if 
two points coincide at the initial moment -- it is clear in advance that 
theirs velocities are infinite at this moment.} The equation
\[{
\ddot{r}(0) = - 4\pi r^2(0)~dp/d\mu
}\]
governs the behavior of the cavity -- a matter on its surface exists 
solely as a shock wave which is singular as $r(0)$$\rightarrow$0. 
Inertial matter can not survive physical conditions nearby the moment of 
the self-collision. In the conceptual framework of modern physics, 
central singularity of a compressional wave can be interpreted solely in 
terms of a pure emission: the collapse $\dot{r}(0)$$<$0 causes raising 
emission, after-rebound process $\dot{r}(0)$$>$0 causes condensation of 
the emission. Spherically symmetric wave is an acoustic wave at the 
surface of a sphere, however, when it reaches the centre, it goes 
through the centre as the flash of emission. 

On the condition $|\ddot{r}(1)|$$\ll$$\mu/r^2$, eq.\ref{sele} could be 
locally approximated by the HSE-equation for $r$$\approx$1. Until 
seemingly negligible superficial vibration is identified as the 
phenomenon of central origin, there is no perceptional reason to cast 
doubt on the possibility of the ideal HSE-state of the Earth. However, 
there are no negligible vibrations at all - every superficial vibration 
is singular at the centre. Fortunately, the smallness of the superficial 
vibration is the essential condition for a planet to be inhabited.

\section{Elementary Lagrangian pulsar (L-pulsar)}
\label{seq:tp}

The simplest model that illustrates the singularity roughly is the model 
of the ideal fluid sphere of volume $V_0$=$4/3\pi R_0^3$ $\Rightarrow$ 
$\rho$=$M/V_0$=\emph{const}. Evidently $\delta H$=0, and the energy 
conservation is reduced to $K$+$U$=\emph{const}. This equation can be 
reduced to the first-order ODE directly. Since the pressure is not the 
function $p(\rho)$ in this case, the CE is used instead: for arbitrary 
radius
 $x( \mu)$ valid\\
$v( \mu,t)-v(0,t)=\frac{4}{3}\pi (x^3-r_L^3)=\mu/\rho~\Rightarrow$\\
$r_L^2\delta r_L=x^2\delta x~\Rightarrow~
r_L^2{\dot{r}}_L=x^2\dot{x}$; where $r_L(t)$=$r(0,t)$\\
The CE gives an advantage to express current state of a sphere via $r_L$ 
and ${\dot{r}}_L$. Basic functions of the elementary pulsar become
\begin{eqnarray}
&~&K(r_L,{\dot{r}}_L)=2\pi\rho ~{\dot{r}}_L^2 r_L^3 (1-\frac{r_L}{R}) 
\nonumber\\
&~&U(r_L)=-\frac{16}{3}\pi ^2~G\rho ^2~R^5~
[~\frac{1}{5}- \frac{1}{2}{( \frac{r_L}{R})}^3 + \frac{3}{10}{( 
\frac{r_L}{R})}^5~],\nonumber
\end{eqnarray}
where $R(t)$=$r(M,t)$. Let $t$=0 be the moment of the self-collision; 
let $t_c$ be the collapse duration time. Then, the moment $-t_c$ 
corresponds to the apex-point $r_{max}$ of the trajectory: 
$r_L(-t_c)$=$r_{max}$; ${\dot{r}}_L(-t_c)$=0; $K(r_{max},0)$=0, hence 
integral of motion $K$+$U$=$U(r_{max})$ is known. Dimensionless relative 
radius $a$=$r_L/R$ images semi-infinite segment 0$<$$r_L$$<$$\infty$ 
into finite one 0$<$$a$$<$1:
\[{
R(a)=R_0{(1-a^3)}^{-1/3};~
{\dot{r}}_L(a)=R_0\dot{a}{(1-a^3)}^{-4/3}
}\]
With the redefinitions
\[{
r_{max}=a_mR_0(1-a_m^3)^{-1/3};~t=t_g\tau;~t_g=\frac{a_m}{\sqrt{(8/9)\pi 
G\rho}},
}\]
initial relative radius $a_m$ substitutes $r_{max}$ and $\tau 
_c$=$t_c/t_g$ becomes dimensionless collapse duration time. Due to these 
definitions, energy conservation equation is reduced to one-parametric 
Cauchy problem at the initial condition $a(-{\tau}_c)$=$a_m$
\begin{eqnarray}
\dot{a}=\pm a_m\sqrt{\frac{(1-a^3)^{11/3}}{(1-a)}~
\frac{U(a_m)-U(a)}{a^3}}, \nonumber
\end{eqnarray}
\[{
\hbox{where } 
U(a)=-\frac{3}{5}~\frac{1-\frac{5}{2}a^3+\frac{3}{2}a^5}{{(1-a^3)}^{5/3}
}
}\]
is finite [$U(1)$=0] dimensionless potential function of a hollow sphere 
in terms of $a$.
For a small L-pulsar $a_m$$\ll$1, by the approximation $ [ U(a_m)-U(a) ] 
\approx 1/2~(a_m^3-a^3)$ and by the renormalization $\alpha ( \tau)$=$a( 
\tau)/a_m$, the Cauchy problem
\begin{eqnarray}
\dot{\alpha}\approx\pm\sqrt{\frac{1}{2}~\frac{1-\alpha ^3}{\alpha 
^3}},~~\alpha (-{\tau}_c)=1
\label{cpep}
\end{eqnarray}
is obtained. In the vicinity of the self-collision $(\alpha$$\ll$$1)$, 
explicit asymptotic trajectory becomes quite simple\\ $\alpha ( 
\tau)$$\approx$$(25/8)^{1/5~}|\tau|^{~2/5}$. Interestingly, Lord 
Rayleigh \cite{ray17} has obtained the same asymptotic estimation of the 
cavitating bubble $r(t)$$\propto$$|t|^{~2/5}$ long ago.

\begin{figure}
\includegraphics{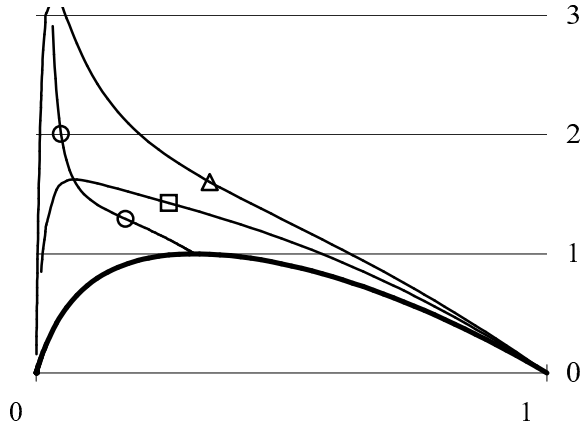}
\caption{Pressure dynamics. Relative profile $p(x,\tau)/p(y,\tau _c)$ 
versus relative depth $(x-r_L)/(R-r_L)$ at $a_m$=$0.1$; solid curve 
$\sim~\tau=\tau _c$; box$~\sim ~\tau$=0.15; $\bigtriangleup~\sim 
~\tau$=0.0655; $\bigcirc\sim~$trajectory of the TPS-pressure.}
\label{fig1}
\end{figure}

As soon as the trajectory is known now, the pressure profile is governed 
by eq.\ref{sele}:
\[{
p(x,\tau)=\rho\int_x^R (\ddot{z}-g)dz,
}\]
\[{
\hbox{where }
\ddot{z}=z^{-2} [ { (r_L^2\dot{r}_L)}'_{\tau}-2z^{-3}(r_L^2\dot{r}_L)^2 
] \hbox{ due to CE }
}\]
Evidently, the top-pressure sphere (TPS)
\[{
y:dp/dz|_{z=y}=(\ddot{z}-g)=0
}\]
exists. The TPS-radius is maximal and the TPS-pressure is minimal at 
$\pm\tau _c$. With the collapse development, TPS-radius vanishes as 
$y$$\approx$$2r_L$, and the TPS-pressure goes to infinity 
(Fig.\ref{fig1}). Radial velocity of the cavity goes to infinity as 
$\dot{r}_L$$\propto$$|\tau|^{-3/5}$; the volumetric velocity stops 
everywhere $x^2\dot{x}$=$r_L^2\dot{r}_L$$\propto$$|\tau|^{1/5}$; total 
kinetic energy is concentrated nearby $r_L$ as the Dirac delta function. 
Indeed, within the geometric segment $(r_L,\beta r_L)$, fraction 
$(1-1/\beta)$ of the total kinetic energy is accumulated 
\[{
\frac{1}{2}\int _{r_L}^{~\beta r_L} {\dot{x}}^2d\mu~\rightarrow 
~(1-1/\beta)(U(a_m)-U(0))
}\]
\emph{E.g.}, the vanishing segment $(r_L,3r_L)$ accumulates $\sim$2/3 of 
the total kinetic energy.

Essentially, the cavity is dynamically stable object. The acceleration 
of the collapse is permanently negative singular function ${\ddot 
r}_L$$\propto$$-1/2~ {a_m}^{-1}{\alpha}^{-4}$. Inertial field acts like 
a fictitious antigravity, since it is aimed only outward; this inertial 
field keeps an empty cavity at the centre. The proportion $1/a_m$ shows 
that the smaller cavity is the greater its fictitious antigravity is.

Fortunately, a body effectively hides the radial beating of the 
L-pulsar. Radius $R(t)$ varies from $R_0$ to about $R_0$+$2/9~a_m g_0 
t_g^2$. Its velocity varies within $R'_t$$\approx$$\pm$$1/3~a_m g_0 
t_g$.

\section{Generality}
\label{seq:gn}

The idea that central pulsatile cavity could cause the emission has been 
suggested \cite{sim98} in connection with striking property of the 
single bubble sonoluminescence (SBSL) effect \cite{gai92}: this bubble 
flashes for $\sim$5$\times$10$^{-11}$~sec at the radius 
$\sim$0.5~$\mu$m. Volumetric squeeze rate exceeds $\sim$10$^5$ for this 
period. Total flux of a flash is $\sim$2$\times 10^{-13}$~J 
\cite{ham01}. Average specific luminosity of the flashing matter is 
$\sim$2$\times$$10^{17}$~W/m$^3$. Its momentary luminosity corresponds 
to the relativistic annihilation of a matter at the rate $\sim$0.3~kg/s 
per cubic meter. This is `a star in a jar' \cite{put98} of $\approx$5~cm 
in diameter. The energetic efficiency of the effect is 
$k$=Flux/$(p\delta v)$$\approx$10$^{-4}$ only. The SBSL effect  serves 
for the adequate laboratory illustration of the L-pulsar. This 
experiment schematically imitates the behavior of a heavenly body: 
liquid sphere imitates inertial and elastic properties of a body, 
central bubble simulates the behavior of the Lagrangian cavity, and the 
driving pressure is an artificial substitute of the effect of 
self-gravitation.

In general features, radial vibrations of an elastic self-colliding 
sphere is quit complex problem. A body contains a tangle of acoustic 
waves which are excited by self-collisions. And vice versa, these waves 
form the radius of the cavity -- the cavity vibrates and collapses 
simultaneously. One could guess, that relatively stable standing waves 
are excited. On the condition of the continuous vibrational spectrum, a 
pulsar is Fourier-undetectable even being visible (fibrillating pulsar). 
On the condition of the strict resonance between the period of the 
gravitational collapse and the period of some prevailing vibrational 
eigenmode, a powerful Fourier-detectable cavity reveals itself as a 
pulsar. However, non-prevailing modes are presented all the time -- 
these are seen as a wide variety of morphologies of a pulse profile.

Evidently, classic model eqs.\ref{sele}-\ref{init} illustrates the 
pulsar behaviour in a general way. The process of the self-collision has 
to be refined in relativistic terms to limit all the infinities of the 
model. Physical conditions within the segment $(r_L,3r_L)$ are 
extraordinary. Figuratively, this is a star turned inside out; its 
concave surface $r_L$ consists of almost strip plasma which exists in 
form of a shock wave; periodic acceleration reproduces bremsstrahlung 
processes in it. These properties give to the observer an illusion of 
super gravity of the super dense convex sphere -- he assumes that he 
observes convex star. By the indirect measurements of the gravity -- 
this cavity is never visible plainly, -- its `free fall acceleration' at 
the radius $r_L$ indicates the fictitious density $\rho 
_{fict}$$\approx$$\rho/a_m^2$. Further, singular velocity causes 
extraordinary Doppler effect. Three factors in total influence upon the 
the spectrum of the emission: (i) dynamical nature of the pulsation; 
(ii) spectral opacity of a body; (iii) Doppler effect. Inwardly, all 
pulsars are similar. Spectral opacity of a sphere-container 
individualizes the spectrum of each pulsar. At the moment of the 
self-collision, energy of the pulsar $[U(a_m)$-$U(0)]$ is concentrated 
at the central point as the clot of emission -- that is a small Big 
Bang. The vast majority of this flash is condensed backward into a 
matter after rebound; some fraction of the flash heats up the 
body-container; a tiny fraction burst through the body. Every pulsar is 
$\gamma$-emitter.

Theoretically, singular inertial field $\ddot{r}_L$ could cause radial 
redistribution of the electric charge due to inertial charge separation: 
negative pulsar - positive body. Electric charge of the cavity limits 
the singularity of self-collision. For a small L-pulsar $a_m$$\ll$1 with 
a small dimensionless charge $\varepsilon$=$2q^2/(Ga_m^4M^2)$; 
$\varepsilon$$\ll$1, the Cauchy problem (\ref{cpep}) can be approximated 
roughly as
\[{
\dot{\alpha}\approx\pm\sqrt{1/2~ 
[-\varepsilon{\alpha}^{-4}+(1+\varepsilon){\alpha}^{-3}-1~] },~~\alpha 
(-{\tau}_c)=1
}\]
Then, finite rebound $\dot{\alpha}$=0 occurs at 
$\alpha$$\approx$$\varepsilon/(1+\varepsilon)$. Interestingly, quite 
moderate axial rotation of this configuration yields another 
understanding of the source of the magnetic field of a body and of a 
pulsar.  

The pulsar of the Earth is actual geophysical problem. Perceptionally, 
it is very weak pulsar. It excites free oscillations of the Earth 
(\emph{e.g.} \cite{kuz98}), however these are generatrix oscillations 
$\sim$1$\div$10~mHz. To estimate its parameters, let the pulsar provide 
annual geothermal energy output $\sim$10$^{21}$~J at the SBSL efficiency 
$k$=10$^{-4}$ -- \emph{i.e.} the pulsar burns down the Earth's mass at 
the rate $ka_m^3GM^2/(2Rc^2)$ per cycle. Then, for 
$\rho$$\approx$5500~kg/m$^3$; $R_0$$\approx$6.4$\times$10$^6$~m, 
corresponding cavity $a_m$$\approx$3$\times$10$^{-6}$ 
($r_{max}$$\approx$20~m) pulsates at average frequency $\sim$150~Hz; 
superficial amplitude is $\approx $3$\times$10$^{-10}$~m; superficial 
velocity varies in $\approx$$\pm$6$\times$10$^{-8}$~m/s. The pulsar 
generates faint noise-like acoustic field which constitutes some part of 
the geoacoustic ambient noise. Minimal fictitious `free fall 
acceleration' on the surface of this `neutron star' is 
$\ddot{r}_L$$\approx$1.5$\times$10$^5g_0$ and corresponding minimal 
fictitious density is $\rho _{fict}$$\approx$10$^{14}$~kg/m$^3$.

\section{Discussions}
\label{seq:ds}

There are two discussible points of non-HSE pulsar: (i) the problem of 
the observability of the deepest region of a star; (ii) positive period 
derivative of every pulsar. Direct observability of a pulsar is a 
relative notion: it depends upon its total flux (roughly $\propto a_m^3$ 
-- small but rapidly increasing function) and upon the integral opacity 
of every spectral window of a body. Modern comprehension of the pulsar 
is that it is \emph{Fourier-detectable} phenomenon - otherwise it is not 
identified as a pulsar. To be Fourier-detectable, it must be resonant. 
As being resonant, it is powerful phenomenon. Thus, resonant 
Fourier-detectable high-amplitude Lagrangian cavity has been discovered 
by Bell and Hewish in 1967, and was called `a pulsar'. When L-pulsar is 
resonant, its  thermal output increases, a star is warming up, a pulsar 
is growing up, its period increases as $\propto a_m$. Mostly, a pulsar 
is Fourier-undetectable, however sometimes some pulsar reveals itself as 
the `transient source' of emission. This is the result of the phase 
coincidence of several neighbouring vibrational eigenmodes -- 
\emph{i.e.} harmonic beating. One fibrillating pulsar can be detected 
nonetheless---the pulsar of the Earth. Regular nature of its noise-like 
acoustic signal can be revealed by the synchronous geoacoustic 
observations at several distant geosites. 

\section{Conclusions}
\label{seq:conc}
Classic Euler-Lagrange equation of hydrodynamic flow of radially 
symmetric spheric structure is basically singular one at the central 
point. The state of hydrostatic equilibrium of self-gravitating sphere 
is impossible due to this singularity; a sphere vibrates radially. 
Pulsatile Lagrangian cavity is the physical bearer of the central 
singularity; it provides the  normalizable instantaneous transmission of 
the finite-energy vibrational wave through the central point in the form 
of the flash of emission. This cavity is not a theoretical invention -- 
this one is an inherent feature of the Lagrangian definition of the mass 
configuration of a system. Repeatable \emph{process} of the reversible 
collapse of this cavity is a pulsar. 

\end{document}